\documentclass{particles2015}
\usepackage{graphicx}
\graphicspath{{fig/}}
\usepackage{amsmath}
\usepackage{amsfonts}
\usepackage{amssymb}
\usepackage{bm}
\usepackage{color}
\title{A Master equation for force distributions in polydisperse frictional particles}
\author{Kuniyasu Saitoh$^1$, Vanessa Magnanimo$^1$, and Stefan Luding$^1$}
\heading{Kuniyasu Saitoh, Vanessa Magnanimo and Stefan Luding}
\address{$^{1}$ Faculty of Engineering Technology, MESA+, University of Twente\\
Drienerlolaan 5, 7522 NB, Enschede, The Netherlands}
\abstract{
An incremental evolution equation,\ i.e.\ a Master equation in statistical mechanics, is introduced for force distributions in polydisperse frictional particle packings.
As basic ingredients of the Master equation, the conditional probability distributions of particle overlaps are determined by molecular dynamics simulations.
Interestingly, tails of the distributions become much narrower in the case of frictional particles than frictionless particles,
implying that correlations of overlaps are strongly reduced by microscopic friction.
Comparing different size distributions, we find that the tails are wider for the wider distribution.
}
\begin{document}
%
%
%
%
%
\section{INTRODUCTION}
\label{sec:intro}
Quasi-static deformations of soft particles,\ e.g.\ glasses, colloids, and granular materials, have been widely investigated because of their importance in industry and science.
However, their macroscopic behaviors are still not fully understood due to disordered configurations and complex dynamics \cite{lemaitre}.
At the microscopic scale, their mechanical response is probed as a reconstruction of the force-chain network \cite{gn2},
where non-affine displacements of particles cause a complicated restructuring of the network including also opening and closing contacts.
If a macroscopic quantity,\ e.g.\ stress tensor, is defined as a statistical average in the force-chain,
its response to quasi-static deformations is governed by the change of probability distribution function (PDF) of forces.
Therefore, the PDFs in soft particle packings have practical importance such that many theoretical studies \cite{ft4,en4}
have been devoted to determine their functional forms observed in experiments \cite{ps1} and simulations \cite{pd3}.

Recently, we have proposed a Master equation for the PDFs as a stochastic approach towards
a microscopic theory for quasi-static deformations of two-dimensional bidisperse frictionless particles \cite{saitoh2}.
The Master equation can reproduce stochastic evolution of the PDFs under isotropic (de)compressions,
where the conditional probability distributions (CPDs) for the Master equation fully encompass the statistics of microscopic changes of force-chain networks.
In addition, any changes of macroscopic quantities constructed from the moments of forces can be predicted by the Master equation.

In this paper, we generalize our stochastic approach towards wider four-disperse size distributions and include friction in the contacts.
We determine the CPDs from MD simulations and compare them with the cases of polydisperse frictionless and bidisperse frictionless particles.
First, we explain our MD simulations in Sec.\ \ref{sec:method}.
Then, we show our results in Sec.\ \ref{sec:result} and conclude in Sec.\ \ref{sec:summary}.
%
\section{METHOD}
\label{sec:method}
%
We use MD simulations of two-dimensional four-disperse mixtures of frictional soft particles.
The number of constituents ($N_1$, $N_2$, $N_3$, and $N_4$) and the size distribution
are the same with those used in the experiments of wooden cylinders \cite{experiment} as listed in Table \ref{tab:setup},
where the mass, $m$, is identical in simulations to be used for the unit of mass.
The normal force between the particles in contact is given by $f_{ij}^\mathrm{n}=k_\mathrm{n}x_{ij}-\eta_\mathrm{n}\dot{x}_{ij}$
with a normal stiffness, $k_\mathrm{n}$, and normal viscosity coefficient, $\eta_\mathrm{n}$.
Here, an overlap between the particles $i$ and $j$ is introduced as
\vspace{0cm}
\begin{equation}
x_{ij} = R_i+R_j-d_{ij}
\label{eq:xij}
\end{equation}
with an interparticle distance, $d_{ij}$, and the particles' radii, $R_i$ and $R_j$.
Thus, the relative speed in the normal direction is given by its time derivative, $\dot{x}_{ij}$.
The tangential force is introduced as $f_{ij}^\mathrm{t}=k_\mathrm{t}y_{ij}-\eta_\mathrm{t}\dot{y}_{ij}$ which is switched to the sliding friction,
$f_{ij}^\mathrm{s}=-\mu|f_{ij}^\mathrm{n}|$, when the tangential force exceeds the critical value,\ i.e.\ if $|f_{ij}^\mathrm{t}|>\mu|f_{ij}^\mathrm{n}|$,
where $k_\mathrm{t}=k_\mathrm{n}/2$, $\eta_\mathrm{t}=\eta_\mathrm{n}/4$, and $\mu$ are a tangential stiffness, tangential viscosity coefficient, and friction coefficient, respectively.
Here, $y_{ij}$ and $\dot{y}_{ij}$ are a relative displacement and speed in the tangential direction, respectively (see the details in Ref.\ \cite{dem}).

To make static packings of $N(=N_1+N_2+N_3+N_4)=1872$ particles, we randomly distribute them in a $L\times L$ square periodic box,
where no particle touches others and the friction coefficient is set to zero, $\mu=0$ (i.e.\ we use frictionless particles during the preparation of static packings).
Then, we rescale every radius as $R_i(t+\delta t)=\left[1+\left\{\bar{x}-x_\mathrm{m}(t)\right\}/\lambda\right]R_i(t)$~$(i=1,\dots,N)$,
where $t$, $\delta t$, $\bar{x}$, and $x_\mathrm{m}(t)$ are time, an increment of time, a target value of averaged overlap, and the averaged overlap at time $t$, respectively.
Here, we use a long length scale $\lambda=10^2\bar{\sigma}$ with the mean diameter in the final state, $\bar{\sigma}$, to rescale each radius gently
\footnote{We confirmed that static packings prepared with longer length scales, $\lambda=10^3\bar{\sigma}$ and $10^4\bar{\sigma}$, give the same results
concerning their power-law behaviors,\ i.e.\ \emph{critical scaling}, near jamming \cite{gn3}, while we cannot obtain the same results with a shorter length scale, $\lambda=10\bar{\sigma}$.}.
During the rescaling, each radius increases if the averaged overlap is smaller than the target value,
$x_\mathrm{m}(t)<\bar{x}$, and vice versa, so that the averaged overlap will finally converge to $\bar{x}$.
Note that neither particle masses nor the ratios between different diameters change during the rescaling,\ i.e. $\sigma_i(t+\delta t)/\sigma_j(t+\delta t)=\sigma_i(t)/\sigma_j(t)$.
We stop the rescaling when every acceleration of particles drops below a threshold, $10^{-6}k_\mathrm{n}\bar{\sigma}/m$, and assume that the system is static.

\begin{figure}[t]
\centering
\includegraphics[width=\columnwidth]{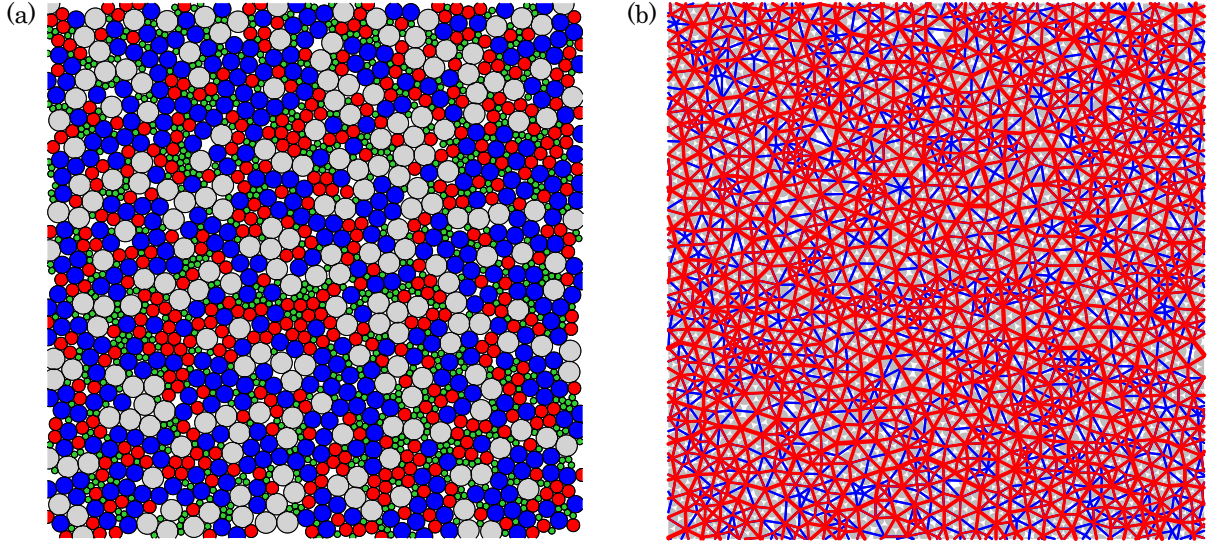}
\caption{(Color online)
(a) A static particle packing with an averaged overlap, $\bar{x}=3\times10^{-6}\bar{\sigma}$, where the total number of particles is $N=1872$.
(b) The complete Delaunay triangulation (DT) of the static packing in (a):
The red solid lines are equivalent to force-chains, where their widths are proportional to the magnitudes of contact forces.
The blue solid lines connect the nearest neighbors without contacts.
The gray circles represent the particles.
\label{fig:pack}}
\end{figure}

\begin{figure}[t]
\centering
\includegraphics[width=13cm]{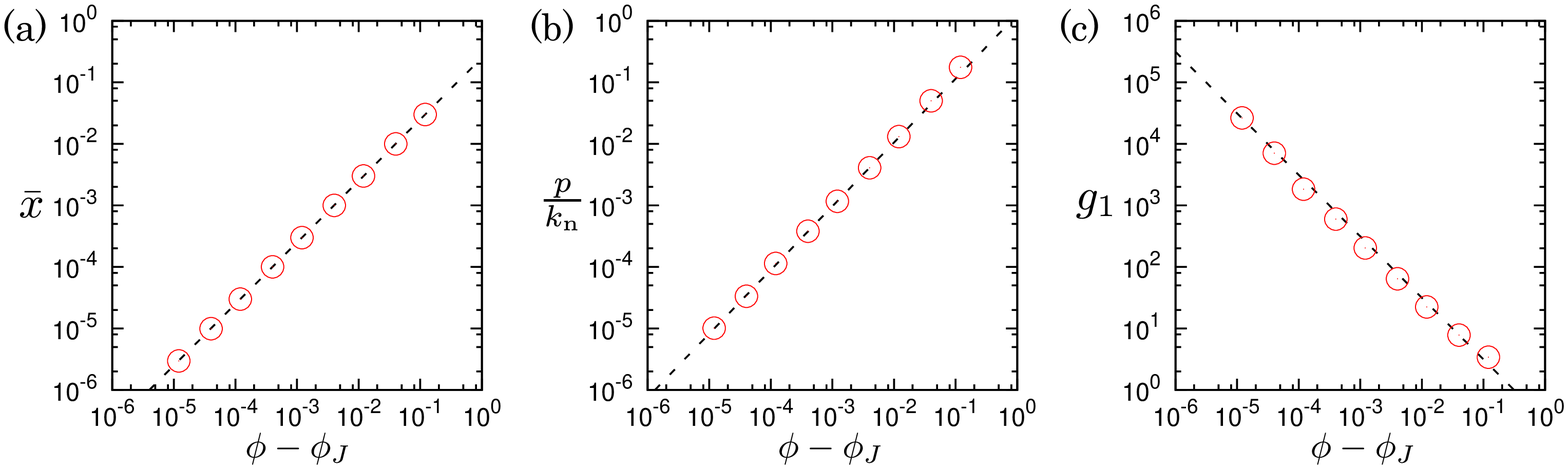}%
\caption{(Color online)
Double logarithmic plots of the (a) averaged overlap, $\bar{x}$, (b) static pressure scaled by the normal stiffness,
$p/k_\mathrm{n}$, and (c) first peak of radial distribution function for scaled distance, $g_1$,
where the dotted lines are power law fittings, (a) $\bar{x}=0.25\bar{\sigma}(\phi-\phi_J)$,
(b) $p/k_\mathrm{n}=1.25\times(\phi-\phi_J)^{1.04}$, and (c) $g_1=0.31(\phi-\phi_J)^{-1}$, respectively.
\label{fig:pss}}
\end{figure}

Figure \ref{fig:pack}(a) is a snapshot of our simulation, where the system is static with the averaged overlap,
$\bar{x}=3\times10^{-6}\bar{\sigma}$, and each color corresponds to each constituent as listed in Table \ref{tab:setup}.
Figure \ref{fig:pack}(b) shows the complete Delaunay triangulation (DT) of the same packing in Fig.\ \ref{fig:pack}(a),
where the red solid lines represent force-chain networks, while the blue solid lines connect the nearest neighbors without contacts.
In our simulations, the distances from jamming are determined by the critical scaling of averaged overlap, $\bar{x}=A(\phi-\phi_J)$ \cite{gn3},
where the critical amplitude is found to be $A\simeq0.25\bar{\sigma}$ as shown in Fig.\ \ref{fig:pss}(a).
We also confirm other critical scalings of the static pressure divided by the normal stiffness (Fig.\ \ref{fig:pss}(b))
and the first peak value in radial distribution function of scaled distance (Fig.\ \ref{fig:pss}(c)),
where we find that the power law dependence on the distance from jamming are given by
$p/k_\mathrm{n}=1.25\times(\phi-\phi_J)^{1.04}$ and $g_1=0.31(\phi-\phi_J)^{-1}$, respectively
\footnote{Here, $p/k_\mathrm{n}$ is dimensionless in two-dimension and the scaled distances in radial distribution functions are defined as $r\equiv d_{ij}/(R_i+R_j)$.} \cite{gn3}.

Then, we switch the friction coefficient to $\mu=0.5$ and apply an isotropic compression to the prepared packings by multiplying every diameter by $\sqrt{1+\delta\phi/\phi}$
such that the area fraction increases from $\phi$ to $\phi+\delta\phi$.
After compression, we relax the system until every acceleration of particles drops below the threshold again.
%
\begin{table}[h!]
\caption{The number of particles per species, $N_s$, diameters, $\sigma_s$, and masses, $m_s$, of the four kinds of particles ($s=1,2,3,4$)
resembling values used in experiments with wooden cylinders in Ref.\ \cite{experiment}.
Each color in Fig.\ \ref{fig:pack} are also listed in the last column.
\label{tab:setup}}
\begin{center}
\begin{tabular}{*{5}{c}}
$s$ & $N_s$ & $\sigma_s/\sigma_4$ & $m_s/m_4$ & colors \\ \hline
$1$ & $807$ & $0.3$ & $1.0$ & green \\
$2$ & $434$ & $0.6$ & $1.0$ & red \\
$3$ & $414$ & $0.8$ & $1.0$ & blue \\
$4$ & $217$ & $1.0$ & $1.0$ & gray \\
\end{tabular}
\end{center}
\end{table}
%
\section{RESULTS}
\label{sec:result}
%
In this section, we introduce a Master equation for the PDFs of forces as a stochastic description of microscopic changes of force-chain networks.
First, we study microscopic responses of force-chain networks to isotropic compressions (Sec.\ \ref{sub:micro}),
where we describe mean values and fluctuations of overlaps in terms of applied strain increments and distances from jamming (Sec.\ \ref{sub:mean}).
We then introduce a Master equation (Sec.\ \ref{sub:master}) and determine transition rates for the Master equation (Sec.\ \ref{sub:cpd}).
\subsection{Microscopic response}
\label{sub:micro}
At the microscopic scale in soft particle packings, the mechanical response to quasi-static deformations is probed as reconstruction of force-chain networks,
where complicated particle rearrangements cause the recombination of force-chains including also opening and closing contacts.
To take into account such changes in structure, we employ the complete Delaunay triangulation (DT) of static packings as shown in Fig.\ \ref{fig:pack}(b),
where not only the particles in contacts, but also the nearest neighbors without contacts,\ i.e.\ \emph{virtual contacts}, are connected by Delaunay edges.
We then generalize the definition of ``overlaps" from Eq.\ (\ref{eq:xij}) to
\vspace{0cm}
\begin{equation}
x_{ij}\equiv R_i+R_j-D_{ij}~,
\label{eq:go}
\end{equation}
where $D_{ij}$ is the Delaunay edge length and the overlaps between particles in virtual contacts ($R_i+R_j<D_{ij}$) are defined as negative values.
Because the DT is unique for each packing, contacts and virtual contacts are uniquely determined.

If we apply an isotropic \emph{affine} compression to the system, every generalized overlap, Eq.\ (\ref{eq:go}),
(not only contacts, but also virtual contacts) changes to
\vspace{0cm}
\begin{equation}
x_{ij}^\mathrm{affine} = x_{ij}+\frac{D_{ij}}{2\phi}\delta\phi~,
\label{eq:xij_affine}
\end{equation}
where we neglected the higher order terms
proportional to $x_{ij}\delta\phi$ and $\delta\phi^2$.
However, the particles are randomly arranged and the force balance is broken for each particle by the affine deformation.
Therefore, the system relaxes to a new static state, where non-affine displacements of particles cause complex changes of contacts including opening and closing contacts.
After the relaxation, overlaps change to new values, $x'_{ij}\neq x_{ij}^\mathrm{affine}$, that is \emph{non-affine responses} of overlaps.
%
\begin{figure}[t]
\centering
\includegraphics[width=\columnwidth]{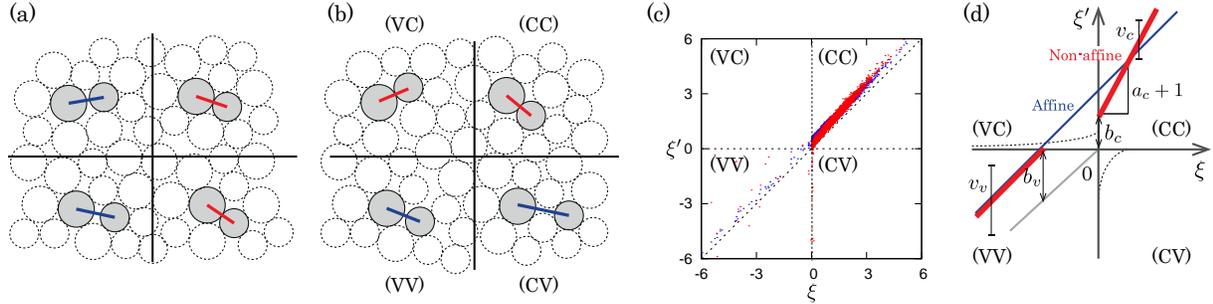}%
\caption{(Color online)
(a) and (b): Sketches of static packings (a) before and (b) after (de)compression, where the red and blue solid lines represent contacts and virtual contacts, respectively.
The four kinds of transitions, (CC) contact-to-contact, (VV) virtual-to-virtual, (CV) contact-to-virtual, and (VC) virtual-to-contact, are displayed.
(c) A scatter plot of scaled overlaps, where the red and blue dots are $\xi'$ and $\xi^\mathrm{affine}$ plotted against $\xi$, respectively.
(d) A schematic picture of affine and non-affine responses of scaled overlaps,
where the blue and red solid lines represent $\xi^\mathrm{affine}$ (in average) and the linear functions, $m_s(\xi)$, respectively.
The slope in (CC), $a_c$, and all the dimensionless lengths, $b_s$ and $v_s$, are proportional to $\gamma$, while the slope in (VV) is negligible, $a_v\simeq0$.
\label{fig:scat}}
\end{figure}

As shown in Figs.\ \ref{fig:scat}(a) and (b), there are only four kinds of transitions from $x_{ij}$ to $x'_{ij}$:
A positive overlap, $x_{ij}>0$, remains as positive, $x'_{ij}>0$, or a negative overlap, $x_{ij}<0$, stays in negative, $x'_{ij}<0$,
where they do not change their signs and thus contacts are neither generated nor broken.
We call these changes ``\emph{contact-to-contact} (CC)" and ``\emph{virtual-to-virtual} (VV)", respectively.
The other cases are that a positive overlap changes to a negative one and a negative overlap becomes positive,
where existing contacts are broken and new contacts are generated, respectively.
We call these changes opening and closing contacts, or in analogy to the above, ``\emph{contact-to-virtual} (CV)" and ``\emph{virtual-to-contact} (VC)", respectively.

In the following, we scale the generalized overlaps by the averaged overlap before compression
such that scaled overlaps before compression, after affine deformation, and after relaxation are introduced as
$\xi\equiv x_{ij}/\bar{x}$, $\xi^\mathrm{affine}\equiv x^\mathrm{affine}_{ij}/\bar{x}$, and $\xi'\equiv x'_{ij}/\bar{x}$, respectively (we omit the subscript, $ij$, after the scaling).
From Eq.\ (\ref{eq:xij_affine}) and the critical scaling, $\bar{x}=A(\phi-\phi_J)$, the scaled overlap after affine deformation is found to be
\vspace{0cm}
\begin{equation}
\xi^\mathrm{affine}=\xi+B_a\gamma
\label{eq:xi_affine}
\end{equation}
which is a \emph{linear function} of $\xi$ with an offset, $B_a\gamma\equiv \left(D_{ij}/2A\phi\right)\gamma$,
proportional to the scaled strain increment, $\gamma\equiv\delta\phi/\left(\phi-\phi_J\right)$.
However, the scaled overlap after non-affine deformation, $\xi'$, fluctuates around mean due to complicated particle rearrangements during the relaxation.
\subsection{Mean and fluctuation}
\label{sub:mean}
To describe scaled overlaps after non-affine deformation, $\xi'$, we measure their mean and fluctuations through scatter plots.
Figure \ref{fig:scat}(c) displays the scatter plot, where the four different transitions under compression are mapped onto four regions:
(CC) $\xi,\xi'>0$, (VV) $\xi,\xi'<0$, (CV) $\xi>0$, $\xi'<0$, and (VC) $\xi<0$, $\xi'>0$, respectively.
In this figure, overlaps after affine deformation are described by the \emph{deterministic} equation (\ref{eq:xi_affine}),
while those after non-affine deformation distribute around \emph{mean values} with finite \emph{fluctuations}.
The differences between affine and non-affine responses are always present, but not visible
if the applied strain is small or the system is far from jamming,\ i.e.\ if $\gamma\ll1$,
while $\xi'$ deviates more from $\xi^\mathrm{affine}$ and data points are more dispersed if $\gamma\gg1$.

\begin{figure}[t]
\centering
\includegraphics[width=13cm]{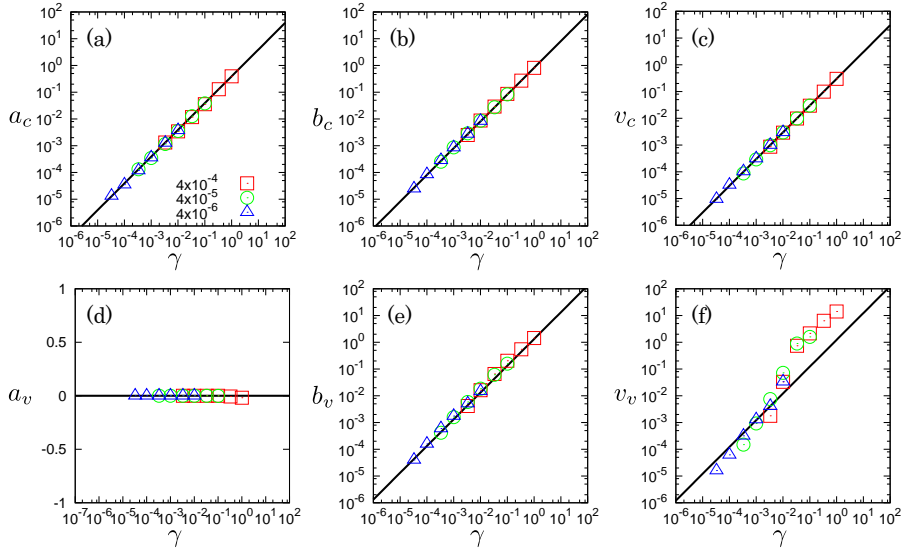}
\caption{(Color online)
Double logarithmic plots of the coefficients for mean values and fluctuations of scaled overlaps as functions of the scaled strain increment,
(a) $a_c$, (b) $b_c$, (c) $v_c$, (e) $b_v$, and (f) $v_v$, and a semi-logarithmic plot of (d) $a_v$,
where we apply strain increments $\delta\phi=4\times10^{-4}$, $4\times10^{-5}$, and $4\times10^{-6}$ (as indicated by the different symbols in the legend of (a))
to the static packings with $\phi-\phi_J=1.2\times10^{-1}$, $4\times10^{-2}$, $1.2\times10^{-2}$, $4\times10^{-3}$, $1.2\times10^{-3}$, and $4\times10^{-4}$.
The solid lines represent the linear scalings,\ Eq.\ (\ref{eq:scaling}), between the fitting range, $10^{-6}\le\gamma\le5\times10^{-3}$.
\label{fig:coe}}
\end{figure}

In the same way as affine response, Eq.\ (\ref{eq:xi_affine}), we describe the mean values of $\xi'$ in (CC) and (VV) regions by linear functions of $\xi$,
\vspace{0cm}
\begin{equation}
m_s(\xi) = (a_s+1)\xi+b_s~,
\label{eq:mzt}
\end{equation}
where the subscripts, $s=c$ and $v$, represent the mean values in (CC) and (VV), respectively.
We also introduce standard deviations of $\xi'$ from their means as $v_s$ (which are almost independent of $\xi$).
Then, the systematic deviation from the affine response is quantified by the coefficients, $a_s$, $b_s$, and $v_s$, as summarized in Fig. \ref{fig:scat}(d).
Note that the affine response, Eq.\ (\ref{eq:xi_affine}), is obtained if $a_s=v_s=0$ and $b_s=B_a\gamma$.
Except for $a_v\simeq0$, all the coefficients \emph{linearly} increase with the scaled strain increment (Fig.\ \ref{fig:coe}),
where all data with a wide variety of $\delta\phi$ and $\phi-\phi_J$ collapse onto linear scalings,
\vspace{0cm}
\begin{equation}
a_s = A_s\gamma~,\hspace{3mm} b_s = B_s\gamma~,\hspace{3mm} v_s = V_s\gamma~,
\label{eq:scaling}
\end{equation}
between the fitting range, $10^{-6}\le\gamma\le5\times10^{-3}$, with the scaling amplitudes, $A_s$, $B_s$, and $V_s$, listed in Table \ref{tab:fpdist}.
We observe that $a_v\simeq0$ and $B_v\approx B_a$($\simeq1.3$ in average) such that virtual contacts almost behave affine in average except for their huge fluctuations ($V_v\gg V_c$).
In contrast, $B_c$ is always smaller than $B_a$ such that $m_c(\xi)$ intersects $\xi^\mathrm{affine}$ at $\xi^\ast=(B_a-B_c)/A_c\simeq1.3$, which is independent of $\gamma$.
This leads to small responses, $\xi'<\xi_\mathrm{affine}$, of small overlaps, $\xi<\xi^\ast$, and vice versa,
implying preferred tangential and hindered normal displacements as a sign of non-affine deformations \cite{rs1}.

In contrast to (CC) and (VV), the data of $\xi'$ in (VC) and (CV) are concentrated in narrow regions (the inside of the dashed lines in Fig.\ \ref{fig:scat}(d)),
whereas $\xi^\mathrm{affine}$ linearly increases with $\xi$ in (VC) and there is no data of $\xi^\mathrm{affine}$ in (CV),\ i.e.\ affine responses do not generate opening contacts.
%
%
\begin{table}[h!]
\caption{Scaling amplitudes in Eq.\ (\ref{eq:scaling}), $A_s$, $B_s$, and $V_s$, for polydisperse frictional particles.
The $q$-indices for the CPDs in (CC) and (VV), where $q_s^\mathrm{BL}$, $q_s^\mathrm{PL}$, and $q_s^\mathrm{PN}$ are the $q$-indices
for bidisperse frictionless particles \cite{saitoh2}, polydisperse frictionless particles, and polydisperse frictional particles, respectively.
\label{tab:fpdist}}
\begin{center}
\begin{tabular}{*{7}{c}}
$s$ & $A_s$ & $B_s$ & $V_s$ & $q_s^\mathrm{BL}$ & $q_s^\mathrm{PL}$ & $q_s^\mathrm{PN}$ \\ \hline
$c$ & $0.39$ & $0.81$ & $0.30$ & $1.13$ & $1.72$ & $1.19$ \\
$v$ & $0.00$ & $1.32$ & $1.22$ & $1.39$ & $1.96$ & $1.09$ \\
\end{tabular}
\end{center}
\end{table}
\subsection{Master equation}
\label{sub:master}
The reconstruction of force-chain networks attributed to the changes, (CC), (VV), (CV), and (VC), is well captured by the PDFs of scaled overlaps.
Here, we introduce the PDF as $P_\phi(\xi)$ with the subscript, $\phi$, representing the area fraction in the system.
Because the total number of Delaunay edges is conserved during deformations, the PDFs are normalized as $\int_{-\infty}^\infty P_\phi(\xi)d\xi = 1$.

\begin{figure}[t]
\centering
\includegraphics[width=7cm]{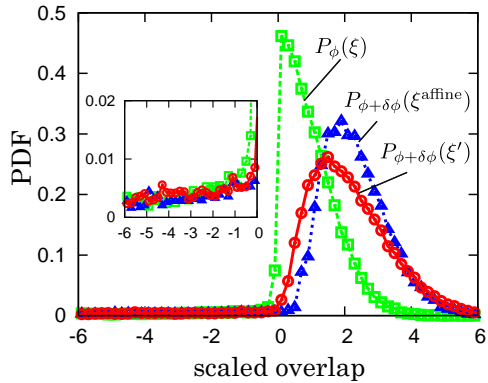}%
\caption{(Color online)
The PDFs of scaled overlaps before compression, $P_\phi(\xi)$ (the squares), after affine deformation,
$P_{\phi+\delta\phi}(\xi^\mathrm{affine})$ (the triangles), and after non-affine deformation, $P_{\phi+\delta\phi}(\xi')$ (the circles),
where the inset is a magnification of the PDFs of negative scaled overlaps.
\label{fig:fpdf}}
\end{figure}

In our previous study of bidisperse frictionless particles \cite{saitoh2},
we found that the affine deformation, Eq.\ (\ref{eq:xi_affine}), just shifted the PDF to the positive direction,
while the non-affine deformation broadened the PDF in positive overlaps and generated a discontinuous ``gap" around zero
\footnote{Such a discontinuity is specific to static packings, where a corresponding gap has been observed in a radial distribution function of glass with zero-temperature \cite{th1}.}.
We also find that the PDF of negative overlaps after non-affine deformation is comparable with that after affine deformation.
In our simulations of polydisperse frictional particles ($\mu=0.5$), we observe similar results to our previous study.
Figure \ref{fig:fpdf} displays the PDFs for polydisperse frictional particles, where the PDF before compression, $P_\phi(\xi)$, has a discontinuous gap around zero.
The affine deformation pushes the PDF towards the positive direction as $P_{\phi+\delta\phi}(\xi^\mathrm{affine})$,
where the discontinuity is smoothed out because of the polydispersity in the system.
After non-affine deformation, the PDF widens in positive overlaps as $P_{\phi+\delta\phi}(\xi')$, while there is no significant difference between
$P_{\phi+\delta\phi}(\xi^\mathrm{affine})$ and $P_{\phi+\delta\phi}(\xi')$ in negative overlaps (the inset in Fig.\ \ref{fig:fpdf}).
%

To describe such the non-affine evolution of the PDFs, we connect the PDF after non-affine deformation to that before compression through the Chapman-Kolmogorov equation \cite{vanKampen},
\vspace{0cm}
\begin{equation}
P_{\phi+\delta\phi}(\xi') = \int_{-\infty}^\infty W(\xi'|\xi)P_\phi(\xi)d\xi~,
\label{eq:chapman-kolmogorov}
\end{equation}
assuming that transitions between overlaps (from $\xi$ to $\xi'$) can be regarded as \emph{Markov processes}.
On the right-hand-side of the Chapman-Kolmogorov equation (\ref{eq:chapman-kolmogorov}),
the conditional probability distribution (CPD) of scaled overlaps, $\xi'$, which were $\xi$ before compression, is introduced as $W(\xi'|\xi)$.
By definition, the CPD is normalized as $\int_{-\infty}^\infty W(\xi'|\xi)d\xi'=1$.
Then, the Master equation for the PDFs is derived from Eq.\ (\ref{eq:chapman-kolmogorov}) as \cite{vanKampen}
\vspace{0cm}
\begin{equation}
\frac{\partial}{\partial\phi}P_\phi(\xi') = \int_{-\infty}^\infty\left\{T(\xi'|\xi)P_\phi(\xi)-T(\xi|\xi')P_\phi(\xi')\right\}d\xi
\label{eq:master}
\end{equation}
with the transition rate defined as $T(\xi'|\xi)=\lim_{\delta\phi\rightarrow0}W(\xi'|\xi)/\delta\phi$.
The first and second terms on the right-hand-side of the Master equation (\ref{eq:master}) represent the gain and loss of new overlaps, $\xi'$, respectively.
Therefore, the transition rates or the CPDs fully determine the statistics of microscopic responses of force-chain networks.
\subsection{Conditional probability distributions}
\label{sub:cpd}
We determine the CPDs of scaled overlaps as the distributions of $\xi'$ around their mean values, $m_s(\xi)$.
For example, the CPD for affine deformation is given by a delta function, $W_\mathrm{affine}(\xi'|\xi)=\delta(\xi'-\xi^\mathrm{affine})$,
which just shifts the PDF by $B_a\gamma$,\ i.e.\ $P_{\phi+\delta\phi}(\xi)=P_\phi(\xi-B_a\gamma)$.
However, non-affine deformations generate fluctuations of scaled overlaps around their mean values so that the CPDs must have finite widths.
In the following, we determine the CPDs for \emph{polydisperse frictionless particles} and those for \emph{polydisperse frictional particles}
to examine the effect of particle friction (characterized by the friction coefficient, $\mu$) on the statistics of microscopic responses of force-chain networks.
We also compare our results with our previous work of \emph{bidisperse frictionless particles} \cite{saitoh2} to study the influence of size distributions.

\begin{figure}[t]
\centering
\includegraphics[width=12cm]{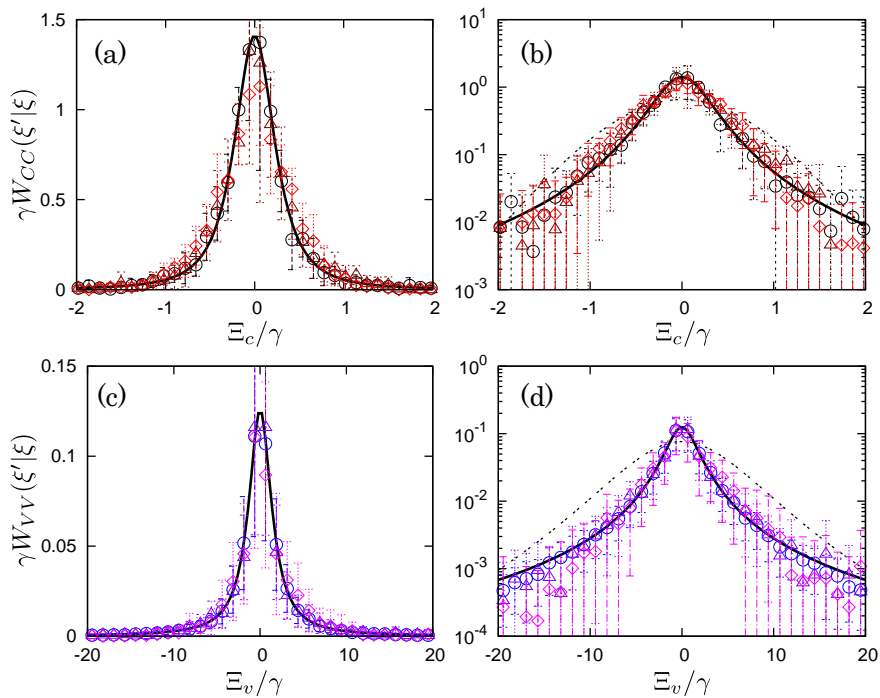}
\caption{(Color online)
The CPDs for \emph{polydisperse frictionless particles} multiplied by $\gamma$ in (a) (CC) and (c) (VV) plotted against scaled distances from mean values, $\Xi_s/\gamma$ ($s=c,v$).
(b) and (d) are semi-logarithmic plots of (a) and (c), respectively.
The solid lines are the q-Gaussian distributions with the $q$-indices, (b) $q_c^\mathrm{PL}=1.19$ and (d) $q_v^\mathrm{PL}=1.09$, respectively.
The dotted lines are the CPDs obtained from our previous study of two-dimensional \emph{bidisperse frictionless particles} \cite{saitoh2},
where the $q$-indices are given by (b) $q_c^\mathrm{BL}=1.13$ and (d) $q_v^\mathrm{BL}=1.39$, respectively.
\label{fig:cpd_zero}}
\end{figure}

Figures \ref{fig:cpd_zero}(a) and (c) show the CPDs for polydisperse frictionless particles ($\mu=0$) in (CC) and (VV), respectively.
As can be seen, all data are \emph{symmetric} around the mean values, $m_s(\xi)$, and are well collapsed
if we multiply the CPDs and the distances from the mean values defined as $\Xi_s\equiv\xi'-m_s(\xi)$ by $\gamma$ and $1/\gamma$, respectively.
In these figures, the solid lines are given by $\gamma W_{CC}(\xi'|\xi)=f_c(\Xi_c/\gamma)$ and $\gamma W_{VV}(\xi'|\xi)=f_v(\Xi_v/\gamma)$
with the \emph{q-Gaussian distribution} \cite{q-gauss0},
\vspace{0cm}
\begin{equation}
f_s(x) = \frac{1}{c(q_s)}\left[1+\frac{x^2}{n(q_s)V_s^2}\right]^{\frac{1}{1-q_s}}~,
\label{eq:q-gauss}
\end{equation}
where the functions are defined as $n(t)=(t-3)/(1-t)$ and $c(t)=V_s\sqrt{n(t)}B\left(1/2,n(t)/2\right)$ with the beta function, $B(x,y)$.
In Table \ref{tab:fpdist}, we list the \emph{q-indices} which characterize shapes of the CPDs.
Note that the $q$-index must be in the range between $1<q_s<3$, where the normal (Gaussian) distribution corresponds to the limit, $q\rightarrow1$.
Figures \ref{fig:cpd_zero}(b) and (d) are semi-logarithmic plots of Fig.\ \ref{fig:cpd_zero}(a) and (c), respectively,
where the dotted lines are the CPDs obtained from our previous study on bidisperse frictionless particles \cite{saitoh2}.
From these results, we observe that the basic properties of CPDs,\ i.e.\ their symmetry and self-similarity for different $\gamma$, are not affected by the size distribution.
However, we find that the CPDs for polydisperse particles have much wider tails than those for bidisperse particles,\ i.e.\ $q_s^\mathrm{BL}\ll q_s^\mathrm{PL}$ in Table \ref{tab:fpdist},
implying that the spatial correlation of scaled overlaps increases with the increase of polydispersity \cite{ogarko}.
%

\begin{figure}[t]
\centering
\includegraphics[width=12cm]{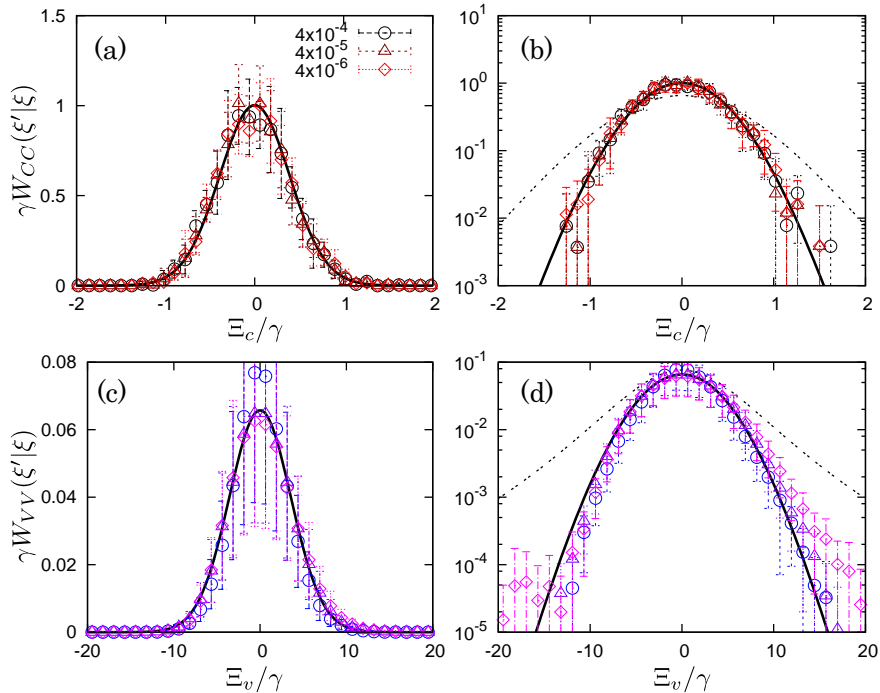}
\caption{(Color online)
The CPDs for \emph{polydisperse frictional particles} multiplied by $\gamma$ in (a) (CC) and (c) (VV) plotted against scaled distances from mean values, $\Xi_s/\gamma$ ($s=c,v$).
(b) and (d) are semi-logarithmic plots of (a) and (c), respectively.
The solid lines are the q-Gaussian distributions with the $q$-indices, $q_c^\mathrm{PN}=1.19$ and $q_v^\mathrm{PN}=1.09$, respectively.
The dotted lines are the CPDs obtained from our previous study of two-dimensional \emph{bidisperse frictionless particles} \cite{saitoh2},
where the $q$-indices are given by (b) $q_c^\mathrm{BL}=1.13$ and (d) $q_v^\mathrm{BL}=1.39$, respectively.
\label{fig:cpd}}
\end{figure}

Figures \ref{fig:cpd}(a) and (c) display the CPDs for polydisperse frictional particles
($\mu=0.5$ with the same polydispersity as in Fig.\ \ref{fig:cpd_zero}) for (CC) and (VV) cases, respectively.
Note that all data are symmetric around their mean values and well collapse after the same scaling as in Fig.\ \ref{fig:cpd_zero}.
Figures \ref{fig:cpd}(b) and (d) are semi-logarithmic plots of Fig.\ \ref{fig:cpd}(a) and (c), respectively,
where the dotted lines are the CPDs for bidisperse frictionless particles \cite{saitoh2}.
Comparing the results in Figs.\ \ref{fig:cpd_zero} and \ref{fig:cpd}, we confirm that the basic properties,\ i.e.\ symmetry and self-similarity, of the CPDs do not change by particle friction,
while the tails of CPDs for frictional particles are much narrower than those for frictionless particles,\ i.e.\ $q_s^\mathrm{PN}\ll q_s^\mathrm{PL}$ in Table \ref{tab:fpdist}.
Therefore, the microscopic friction at the contact drastically decrease the spatial correlations of scaled overlaps.
%
%
\section{SUMMARY}
\label{sec:summary}
%
We have introduced a Master equation for the PDFs of forces in two-dimensional polydisperse frictionless and frictional particle packings.
We find that the microscopic response to isotropic compression increasingly deviates from the affine prediction with the increase of the scaled strain increment, $\gamma$,
in agreement with our previous study of bidisperse frictionless particles \cite{saitoh2}.
The scaling amplitudes for mean values of scaled overlaps are almost the same as those for bidisperse frictionless particles \cite{saitoh2},
implying that the degree of non-affinity depends on neither particle friction nor size distribution in average.
In addition, symmetry and self-similarity of the CPDs, which govern the statistics of microscopic changes of the force-chain network, are not affected by friction and size distribution.
In contrast, the shapes and widths of the CPDs are different for different systems:
By using the $q$-indices to describe their shapes in all cases, we observe that the $q$-index for polydisperse particles is larger than bidisperse particles,
which means that the spatial correlation of scaled overlaps increases with polydispersity (the $q$-index not equal to one means a non-Gaussian distribution).
On the other hand, the $q$-index becomes nearly equal to one if microscopic friction is considered, implying a dramatic decrease of spatial correlation.

In conclusion, the basic properties of the Master equation,\ i.e.\ linear scalings of mean values and fluctuations of scaled overlaps,
and the symmetry and self-similarity of the CPDs, are insensitive to particle friction and size distributions, in the cases tested in this paper,
while the shape of CPDs depends on both friction coefficient and polydispersity, implying different spatial correlations of scaled overlaps for different material properties.
\section*{Acknowledgment}
This work was financially supported by the NWO-STW VICI grant 10828.
A part of numerical computation was carried out at the Yukawa Institute Computer Facility, Kyoto, Japan.
\end{document}